\documentstyle[sevensterapj,sevensterpsfig]{sevensterart}
\def\shortauthors{\lefthead}
\def\shorttitle{\righthead}

\newcount\levelone    \levelone=0
\newcount\leveltwo    \leveltwo=0
\newcount\levelthree  \levelthree=0
\newcount\levelfour   \levelfour=0
\def\chaphead{}                             
\def\secno{\chaphead\the\levelone}
\def\subno{\chaphead\the\levelone.\the\leveltwo}
\def\subsubno{\chaphead\the\levelone.\the\leveltwo.\the\levelthree}
\def\subsubsubno{\chaphead\the\levelone.\the\leveltwo.\the\levelthree
                           .\the\levelfour}
\def\newsec{\advance\levelone by1 \leveltwo=0 \levelthree=0 \levelfour=0}
\def\newsub{\advance\leveltwo by1 \levelthree=0 \levelfour=0}
\def\newsubsub{\advance\levelthree by1 \levelfour=0}
\def\newsubsubsub{\advance\levelfour by1}

\def\absnarrower{\advance\leftskip by \abstractindent}

\newdimen\secskipamount  \secskipamount=1pt
\newdimen\subskipamount  \subskipamount=1pt
\newdimen\bottomtol \bottomtol=0.03\vsize
\def\secskip{\par \ifdim\lastskip<\secskipamount \removelastskip \fi
    \vskip 0pt plus \bottomtol \penalty-250
    \vskip 0pt plus -\bottomtol \relax
    \vskip\secskipamount plus3pt minus3pt}
\def\subskip{\par \ifdim\lastskip<\subskipamount \removelastskip \fi
    \vskip 0pt plus 0.5\bottomtol \penalty-150
    \vskip 0pt plus -0.5\bottomtol \relax
    \vskip\subskipamount plus2pt minus2pt}
\def\subsubskip{\par \ifdim\lastskip<\subskipamount \removelastskip \fi
    \vskip 0pt plus 0.5\bottomtol \penalty-150
    \vskip 0pt plus -0.5\bottomtol \relax
    \vskip\subskipamount plus2pt minus2pt \hskip 10pt}

\outer\def\unnumberedsectionbegin #1 #2\par {\secskip \noindent {{\bf  #1}
\dotfill #2}
    \nobreak \vskip 1pt \noindent}

\outer\def\sectionbegin #1 #2\par {\secskip \newsec \noindent {{\bf \secno\  #1}
\dotfill #2}
    \nobreak \vskip 1pt \noindent}
\outer\def\subsectionbegin #1 #2\par {\subskip \newsub {\subno\ {\rm #1} \hfill
#2}
    \nobreak \vskip 1pt \noindent}
\outer\def\subsubsectionbegin #1 #2\par {\subsubskip \newsubsub
    {\subsubno\ {\it #1} \hfill #2}
  \nobreak \vskip 1pt \noindent}
\newcount\eqnumber \eqnumber=0
\def\new{{\rm\chaphead\the\eqnumber}\global\advance\eqnumber by 1}

\newcount\fignumber \fignumber=0
\def\nfig{\chaphead\the\fignumber\global\advance\fignumber by 1}
\def\ntab{\chaphead\the\tabnumber\global\advance\tabnumber by 1}

\newcount\fononum \fononum=0
\def\nfn{\global\advance\fononum by 1}
\def\fonono{\the\fononum}

\def\bck{\hskip-0.35em}
\def\wisk#1{\ifmmode{#1}\else{$#1$}\fi}
\def\etal{{et al.$\,$}}

\def\msyr{\wisk{\,\rm M_\odot\,yr^{-1}}}

\def\twcc{colour--colour diagram}

\def\mum{\wisk{\mu}m}

\let\lsun=\lsol
\let\msun=\msol

\def\degr{\wisk{^{\circ}}}                                

\def\decdeg#1.#2 {\wisk{#1^{\,\rm o}\bck.\,#2}\ }
\def\decmin#1.#2 {\wisk{#1^{\,\prime}\bck.\,#2}\ }

\def\decsec#1.#2 {\wisk{#1^{\prime\prime}\hskip-0.42em.\hskip0.10em#2}\ }

\def\kms{\wisk{\,\rm km\,s^{-1}\,}}                    

\def\oversim#1#2{\lower1.5pt\vbox{\baselineskip0pt \lineskip-0.5pt
     \ialign{$\mathsurround0pt #1\hfil##\hfil$\crcr#2\crcr\sim\crcr}}}
\def\lsim{\wisk{\mathrel{\mathpalette\oversim{<}}}} 

\def\eqref#1{\advance\eqnumber by -#1 \chaphead\the\eqnumber
           \advance\eqnumber by #1 }
\def\?{\eqref{1}}
\def\last{\advance\eqnumber by -1 {\rm\chaphead\the\eqnumber}\advance
     \eqnumber by 1}
\def\eqnam#1{\xdef#1{\chaphead\the\eqnumber}}
\def\appendixbegin#1 #2{\eqnumber=1 \def\chaphead{{#1}}
    \levelone=0\leveltwo=0\levelthree=0\levelfour=0\eqnumber=1\fignumber=1
    \vskip\subskipamount\noindent{\ninepoint\bf Appendix #1\ \ \ #2}
    \vskip\subskipamount\noindent}
\def\noappendixbegin#1 #2{\eqnumber=1 \def\chaphead{{#1} }
    \levelone=0\leveltwo=0\levelthree=0\levelfour=0\eqnumber=1\fignumber=1
    \vskip\subskipamount\noindent{}
    \vskip\subskipamount\noindent}
\def\nfig{\chaphead\the\fignumber\global\advance\fignumber by 1}
\def\anfig{\global\advance\fignumber by 1}

\def\ntab{\chaphead\the\tabnumber\global\advance\tabnumber by 1}
\def\antab{\global\advance\tabnumber by 1}
\def\nfiga#1{\chaphead\the\fignumber{#1}\global\advance\fignumber by 1}
\def\rfig#1{\advance\fignumber by -#1 \chaphead\the\fignumber
            \advance\fignumber by #1}
\def\fignam#1{\xdef#1{\chaphead\the\fignumber}}
\def\tabnam#1{\xdef#1{\chaphead\the\tabnumber}}

\def\aa#1 #2 {, {A\&A,}{ #1, #2} }
\def\aal#1 #2 {, {A\&A,}{ #1, L#2}\ }
\def\aas#1 #2 {, {A\&AS,}{ #1, #2} }
\def\aj#1 #2 {, {AJ, }{#1, #2}\ }
\def\apj#1 #2 {, {ApJ, }{#1, #2}\ }
\def\apjl#1 #2 {, {ApJ, }{#1, L#2 }\ }
\def\apjs#1 #2 {, {ApJS, }{#1, #2}\ }
\def\araa#1 #2 {, {ARA\&A, }{#1, #2}\ }
\def\mnras#1 #2 {, {MNRAS, }{#1, #2}\ }
\def\bargal#1 { {1996, In: {Buta, R., Crocker, D., Elmegreen, B.~(eds.)
      Barred Galaxies, PASPC 91, San Francisco,} p. #1 }}
\def\thrss#1 { {1999, In: Gibson, B., Axelrod, T., Putman, M.~(eds.)
     The Third Stromlo Symposium: The Galactic Halo,
     PASPC 165, San Francisco, p. #1 }}
\def\apnii#1 {, {2000, In: {Kastner J., Soker N., Rappaport S. (eds.)
     Asymmetric Planetary Nebulae II}. ASP Conf.Ser.199   p. #1}}

\newcount\fignumber \fignumber=1
\newcount\eqnumber \eqnumber=1
\newcount\tabnumber \tabnumber=1

\def\Sct{Section\ }
\def\Eqt{Equation\ }
\def\Fg{Fig.~}

\def\EST{1}

\def\SPEC{1}

\def\RILI{3}

\def\DLI{5}
\def\DRI{6}

\def\IRSB{1}
\def\IRSD{2}
\def\MSXD{3}
\def\IRSE{4}
\def\HFC{5}
\def\SED{6}
\def\LAT{7}

\begin{document}
\def\bibitm{\bibitem{}}

\title{OH--selected AGB and post--AGB stellar objects. \\
   II. Blue versus red evolution off the AGB}

\author{Maartje N.~Sevenster\altaffilmark{1}}

\altaffiltext{1}{sevenste@strw.leidenuniv.nl}

\affil{MSSSO/RSAA, Cotter Road, Weston ACT 2611, Australia}

\shortauthors{M.~Sevenster}
\shorttitle{Blue versus red evolution off the AGB}

\begin{abstract}
Using objects found in a systematic survey of the
galactic Plane in the 1612--MHz masing OH line, we discuss in detail
two ``sequences'' of post--AGB evolution,
a red and a blue.
We argue that the red and the blue groups separate by 
initial mass at $M_{\rm i}$=4$\,$\msun , 
based on evolutionary--sequence turn--off colours,
spectral energy distributions, outflow velocities and
scaleheight. The higher--mass (blue) objects
may have earlier AGB termination.
The lower--mass (red) objects undergo very sudden reddening
for IRAS colour $R_{21}$$\sim$1.2; these sources must all undergo
a very similar process at AGB termination. 
The transition colour
corresponds to average initial masses of $\sim$1.7$\,$\msun .
The combined colour $ 2.5\log((S^{\rm msx}_{21}\,S^{\rm iras}_{12})/
(S^{\rm msx}_{8}\,S^{\rm iras}_{25}))$ proves very sensitive to
distinguish lower--mass, early post--AGB objects from sources still
on the AGB and also to distinguish more evolved post--AGB objects from
star--forming regions. 
The high--mass blue objects are the likely precursors of bipolar
planetary nebulae, whereas the low--mass red objects will evolve
into elliptical planetary nebulae.
\end{abstract}


\section{Introduction}

Maser emission in the satellite line of groundstate 
OH at 1612.231 MHz occurs in a variety of stellar objects.
Star--forming regions (SFR), super giants, visible (Miras)
and invisible (OH/IR) oxygen--rich stars at the tip of
the asymptotic giant branch (AGB) and post--AGB transition 
objects or even young planetary nebulae can all show 
maser emission at 1612 MHz. Spectral
shapes and variability are thought to 
differ for the several different types of object, but
in Paper I (Sevenster 2002), the OH spectral properties were found
to vary less clearly with evolutionary state than usually
assumed. New selection criteria for post--AGB objects were
defined, based largely on mid--infrared colours from the
Midcourse Space Experiment (MSX). Most post--AGB objects are distinguished
by very red mid--infrared colours, but there is also
a group of very blue sources with a strong 60--\mum\ excess. 
These sources are not on the oxygen--rich 
AGB evolutionary sequence (see \Sct 2) and are possible post--AGB
sources according to van der Veen \& Habing (1988).

In this paper, this group of objects will
be compared to the standard red post--AGB objects and 
properties of both groups will be discussed.
We will use 56 objects from the same sample 
{\footnote {A self--explanatory archive of all OH and IR data 
for 766 objects can be downloaded from http://www.mso.anu.edu/$\sim$msevenst}}
as in Paper I, 
of OH--masing sources in the 
galactic Plane (Sevenster \etal 1997a,1997b,2001) with
IRAS (12,25,60,100 \mum ), MSX (4,8,12,15,21 \mum )
and 2MASS identifications (J,H,K) taken
from the respective public databases (see Paper I for details). 
A relatively small percentage of IRAS identifications
is caused largely by (cirrus) confusion at the very low latitudes
of our survey ($|b| < 3^{\circ}.25$).
For the same reason, the flux--density 
measurements, especially at 60 \mum\ and 100 \mum,  are likely 
to be somewhat overestimated even if they are not 
flagged as ``upper limits'' in the IRAS data base.
Except for a few likely supergiants, none of
the sources were identified with an optical counterpart.

In \Sct 2, we define the two groups of post--AGB stars
in more detail and, in \Sct 3, discuss some evolutionary 
aspects. Spectral energy distributions are given in \Sct 4 and
compared to classes of post--AGB stars defined previously
in the literature.
In \Sct 5, age, luminosity and galactic distribution of the 
two groups are discussed and put in the context of planetary nebula
morphology. Conclusions are listed in \Sct 6. 

\section{The two post--AGB samples}

All mid--infrared (MIR) colours are defined as
$[{\rm a}-{\rm b}]=2.5\log(S_{\rm b}/S_{\rm a})$ with $S$ 
flux density in Jy and a,b wavelength in \mum .
For IRAS, the usual names $R_{21}\equiv [12-25]$ and 
$R_{32}\equiv [25-60]$ are used, as well as $R43\equiv [100-60]$.
IRAS magnitudes $[12]\equiv 2.5\log (59.5/S_{12})$
and $[25]\equiv 2.5\log (13.4/S_{25})$ are also used.
For more details on the different bands, see Paper I.
From the 2MASS database, J~(1.25\mum ), H~(1.65\mum ) and/or K~(2.17\mum )
magnitudes were obtained
for 194 sources. Colours are used as uncorrected J--H and H--K.

The two groups of post--AGB stars are defined as follows. 
The blue group, called ``LI'' for lying to the left of the
evolutionary sequence in the IRAS diagram (see \Fg\IRSB ),
are selected by $R_{32}$$\,>\,$$-$0.2 and $R_{21}$$\,>\,$0.2, with

\eqnam\EST
$$ R_{32} > -2.15 + 0.35\,\exp[1.5\,R_{21}] 
\eqno(\new) $$

\noindent
the evolutionary sequence according to van der Veen \& Habing (1988).
Van der Veen \& Habing (1988)
show that this region in the IRAS \twcc\ is partly populated by
post--AGB stars and it also has a connection to the bipolar
post--AGB region (Zijlstra \etal2001; Paper I). Further evidence that our
particular objects are post--AGB stars will come from their
spectral energy distributions (\Sct 4.2).
Only the 25 sources with double--peaked OH spectra are used, to
avoid the inclusion of star--forming regions (SFRs) in the 
sample and to be able to determine the outflow velocities.
The LI stars are listed in Table \DLI .

The red group, ``RI'', consists of 31
double--peaked sources with simply $R_{32}$$\,<\,$1.5 and $R_{21}$$\,>\,$1.4.
They are the ``traditional'' post--AGB sources (Bedijn 1987;
van der Veen \& Habing 1988), turned right
from the evolutionary sequence into IRAS region V and beyond (see \Fg\IRSD ).
They are listed in Table \DRI .
The LI sources have 
higher outflow velocities than the RI sources (\Fg\IRSB ).
The latter
in fact stand out as a separate group with outflow velocities between
9 \kms\ and 15 \kms\ (Paper I).
The LI sources do not only have relatively red $R_{32}$ but also
very red $R_{43}$ as seen in \Fg\IRSD . Otherwise, these 
sources are very blue, as will be substantiated in the 
next sections.

\section{Evolution}

\subsection{Colours}

In \Fg\MSXD\ we present MSX colour
diagrams with symbols according to IRAS properties.
Sources that are still close to the IRAS evolutionary
sequence already show changes in MSX colours (`x' in \Fg\MSXD a,c).
The MSX colour $[12-21]$ also increases suddenly for the reddest sources,
but $[12-15]$ is very constant (\Fg\MSXD b).
In \Fg\MSXD d, the combined colour 
   $R_{21}^{ea} = 
  2.5\log((S^{msx}_{21}\,S^{iras}_{12})/(S^{msx}_{8}\,S^{iras}_{25}))$ 
shows a very marked increase at
$R_{21}$=1.2$\pm$0.2 (see also \Fg\MSXD c). 
The jump occurs as soon as sources redden in $[8-12]$ (see \Fg\MSXD a) and
the subsequent increase in $[15-21]$ 
keeps this colour constant with increasing $R_{21}$.
It is also the best MIR diagnostic to separate (OH--masing) SFRs
with $R_{21}^{ea} < 0.9$ from (evolved) 
post--AGB objects with $R_{21}^{ea} > 0.9$ (\Fg\MSXD d). 
In other infrared colours, these two groups of sources 
show considerable overlap (see Paper I).

It seems likely that the sources at $R_{21}$$\sim$1.2 with high
$R_{21}^{ea}$ are the earliest post--AGB stars (cf.~Paper I).
The limits in $R_{21}$ of this sudden transition correspond to
$R_{32}$=$-$0.6 and $R_{32}$=0.71 on the evolutionary sequence 
(\Eqt\EST ). It is thought that after the AGB, 
the ``standard'' (RI)
sources evolve at fairly constant $R_{32}$ toward redder $R_{21}$,
as shown by Bedijn (1987).
For higher mass--loss rate (Bedijn 1987) and thus higher 
initial mass (Likkel 1989; Garcia--Lario \etal 1993), the
turn--off $R_{32}$ is higher.
The $R_{32}$ limits mentioned above translate
into $M_{\rm i}$$\sim$1\msun\ and  
$M_{\rm i}$$\sim$4\msun\ ($<\,M_{\rm i}\,>$=1.7 \msun\ for an 
IMF with $\alpha$=2.5), following Garcia--Lario \etal(1993).
Now evolved to higher $R_{21}$, the RI sources would all come
from this same mass range.

The LI sources do not fit in this scenario, nor do they
show the colour jump in \Fg\MSXD d.
Rather, they are even bluer than
the evolutionary--sequence sources in some colours.
If these sources are indeed post--AGB sources (\Sct 2), they
must evolve in a very different way from the RI sources.
They have average $[15-21]$=0.1 (0.2 for full sample, 1.1 for RI; \Fg\MSXD (a))
but $[12-15]$=0.4 (0.3 and 0.9; \Fg\MSXD (b)).
This 15--\mum\ excess can be understood in terms of a spectrum
with blue continuum and strong silicate
absorption at 10--\mum\ as well as 18--\mum .
Comparing to the recent assesment of
low--resolution spectra by Chen \etal(2001), we find that in the LI region
of the IRAS diagram
indeed spectra are mostly of their type ``A'', which is the same
as the traditional type ``3n'' of
the low--resolution spectral associations (see IRAS Explanatory Supplement).
This indicates high mass--loss rates and thus high 
initial masses (eg.~Bedijn 1987).
More discussion of the spectral energy distribution will follow
in \Sct 4.

\subsection{Model tracks}

A number of post--AGB evolution 
tracks can be found in the literature. Unfortunately,
a lot of assumptions still have to be made, most significantly
about the evolution of the outflow
velocity and the exact time of the ``end'' of the AGB.
These quantities are frequently kept constant, 
while varying other parameters, such as initial mass.
Van Hoof \etal 1997 (HOW97), however, vary those 
quantities for a range of models with initial mass 3$\,$\msun .
Some of their tracks are shown in the IRAS two--colour diagram (\Fg\IRSE )
and in the combined near--mid--infrared two--colour diagram (\Fg\HFC ; HOW97).

The dashed curves in both figures are for models with ($+$)
silicate dust formation in the post--AGB outflow,
the solid curves for models without($-$). The numbers indicate the 
post--AGB outflow velocity assumed in the corresponding model (15 \kms\ or
150 \kms ; the preceding AGB outflow velocity is 15 \kms\ for both).
The third variable is AGB termination : either ``early'' (at 
period 125 days) or ``standard'' 
(at period 100 days, see HOW97 for details). 
Hence we have models 15$-$,15+,150+,150+(early), etc.

Both post--AGB dust formation and early AGB termination cause
sources to ``loop'' back to very blue $R_{21}$ before resuming
steady  evolution toward higher $R_{21}$.
The 15+ model loops into region VIII (see \Fg\IRSD ), but not to the left of the
evolutionary sequence. To reach such low $R_{21}$ during the
post--AGB evolution, it appears that early AGB termination
is a must. Although HOW97 did not give this model, by extrapolation
it seems that the 15+(early) track could possibly 
explain the IRAS colours of the
LI sources. The same holds for the colours in the
combined infrared diagram (\Fg\HFC ), where the  15+(early) track
would dip down at fairly constant K$-[12]$ to where the few
LI sources with available NIR colours can be found. 
The evolution around the ``turn--around''
tip of the tracks may be rather slow, so relatively many sources
would be found with those colours. 

It should be noted that Steffen \etal(1998) also present models
that ``loop'' through the IRAS two--colour diagram. These are 
evolutionary models for AGB stars in the thermally pulsing phase,
however, that undergo episodes of interrupted mass loss (Zijlstra \etal1992).
Those objects have similar 60--\mum\ excess to our LI stars, but much
bluer $R_{21}$, and are located to the left of region VIb (\Fg\IRSD).

Given our findings in Paper I, it
seems probable that the outflow velocity
does not increase significantly
during the initial post--AGB phase -- as long
as 1612--MHz masers are present -- and that
the 15--\kms\ models are the proper ones for both RI and LI sources.
It is hard to draw any conclusions about the importance of
dust formation in the post--AGB outflow in the absence of models
for a proper range of initial masses, but there is clear 
evidence that both the standard and the early AGB 
termination models may be realistic.
The discussion in \Sct 3.1 would favour the 15$-$ model 
for the RI sources, with continuous $R_{21}$ reddening at 
fairly constant $R_{32}$, and more evidence for this
will be presented in the \Sct 4.1.

For the values of $R_{21}$ out to which we see the OH masers, the HOW97 models
suggest that OH masers might disappear 
around $T_{\rm eff}$=6500$\,$K, both for the RI sources (standard
termination) and for the LI sources (early). For the standard models,
this temperature is reached within 500 yr of leaving the AGB, 
for the early models only after 5000 yr. Hence, for these particular
models of AGB termination, the LI sources would spend ten times
longer being OH--masing post--AGB objects than the RI sources.

\begin{deluxetable}{crrrrrrrrrrr}
\tablecaption{Details for post--AGB classes
I and IVa from van der Veen \etal(1989; see \Sct 4.2). The \.M$_p$ stands
for present post--AGB mass loss and O for the fraction of
oxygen--rich sources in both classes.}
\tablewidth{0pt}
\tablehead{
\colhead{  } &
\colhead{ M$_i$ } &
\colhead{ M$_c$ } &
\colhead{ L } &
\colhead{ h$_z$ } &
\colhead{ O } &
\colhead{log(\.M$_p)$ }
}
\startdata
Class I & 4 & 0.8 & 15000 & 85 &  60 & -5 \\
Class IVa & 1.7 & 0.6 & 5000 & 250  & 80 & -7 \\
\enddata
\end{deluxetable}

\section{Spectral energy distribution}

The spectral energy distributions (SED) are shown for
the RI and LI sources with the most complete spectral 
coverage by IRAS, MSX and 2MASS (\Fg\SED ). For reference, two sources
on the AGB evolutionary sequence are shown as well.
The location of the sources in the IRAS two--colour diagram 
and their outflow velocity are
shown in the insert. The outflow velocities
of the selected LI and RI sources are as expected (\Fg\IRSB ).

The first thing to notice in \Fg\SED\ is that 
the ratio of the IRAS 12--\mum\ to the MSX 12--\mum\ flux density
gives an indication of whether 
the 9.7--\mum\ silicate feature is in emission (bluer AGB source)
or in absorption (redder AGB source); indeed this is as typically
expected.
The SED of the LI sources is very similar to those of the AGB 
sources, but has even redder NIR distribution and
a strong 60--\mum\ excess as noted in \Sct 3.1. Also, they have a
weak 15--\mum\ excess, compared to the AGB sources, and
the silicate feature seems to be in absorption.
The RI sources clearly 
stand out with bluer NIR and, by definition,
redder distributions between 10 and 25 \mum .
The silicate feature appears to be mostly in emission.

\subsection {Comparison to model spectra}

Comparing the SEDs to the model spectra by Bedijn (1987), 
the two AGB sources (\#1\& \#2) have mass--loss
rates of a few times 10$^{-6}$ \msyr .
For the LI sources (\#3\& \#4) no suitable model is
found, since Bedijn did not consider that region of the IRAS two--colour
diagram, but for the RI sources, the post--AGB 
models offer a good explanation.
The NIR colours should correlate strongly with the time since
AGB termination, with the colours becoming bluer with time. The
NIR--to--MIR flux--density ratio increases faster for
sources with lower AGB mass--loss rates and thus lower initial
masses. 

\begin{deluxetable}{crrrrrrrrrrr}
\tablecaption{Details for the sources in \Fg\SED. The $S_{\rm OH}$ is
for the strongest 1612--MHz peak, $S_{25}$ from IRAS.}
\tablewidth{0pt}
\tablehead{
\colhead{ \#  } &
\colhead{ Name (OH$\ell,b$)} &
\colhead{ IRAS } &
\colhead{ $V_{\rm exp}\rm ({km\over s})$ } &
\colhead{ $S_{\rm OH}$(Jy) } &
\colhead{ $S_{25}$(Jy) } &
\colhead{ K }
}
\startdata
1 & OH355.588$-$02.978  & 17434$-$3414 &  6.6 &  0.44 & 41.91 & 4.82 \\
2 & OH350.982$-$02.391  & 17288$-$3748 & 14.6 &  1.38 & 26.63 & 5.04 \\
3 & OH344.929+00.014    & 17004$-$4119 & 19.0 & 39.93 & 322.60 & -- \\
4 & OH010.076$-$00.095  & 18052$-$2016 & 28.5 &  2.02 & 32.45 & 7.57 \\
5 & OH355.641$-$01.742  & 17385$-$3332 & 10.2 &  4.03 & 13.25 & 9.07 \\
6 & OH359.140+01.137    & 17359$-$2902 &  9.5 &  1.66 & 12.38 & 10.40 \\
7 & OH001.212+01.257    & 17404$-$2713 & 13.1 &  4.48 & 20.74 & 11.08 \\
8 & OH353.945$-$00.972  & 17310$-$3432 & 11.7 &  1.39 & 10.85  & 11.47 \\
9 & OH353.973+02.727    & 17164$-$3226 & 13.9 &  2.33 &  9.45 & 12.78 \\
10 & OH007.961+01.445   & 17550$-$2120 & 12.4 &  4.30 & 21.08 & 13.22 \\
\enddata
\end{deluxetable}

From the comparison between the spectral energy distributions
and Bedijn's models, we
estimate that the RI sources (\#5 -- \#10) plotted in \Fg\SED\ had
AGB mass--loss rates of 10$^{-4}$ to 10$^{-3}$ \msyr\ and all left the
AGB less than $\sim$500 yr ago.
The RI sources \#9 and \#10 already have strong 
NIR -- very blue J$-$K -- but still low NIR--to--MIR flux--density ratio.
This suits the models with higher mass loss (10$^{-3}$ \msyr ),
in agreement with the idea that higher--mass sources
have larger turn--off $R_{32}$ (\Sct 3.1).
It is also in agreement with the idea that sources evolve off 
the AGB toward redder $R_{21}$ at almost constant $R_{32}$.

\subsection {Comparison to observed spectra}

Van der Veen \etal (1989) give 5 observational classes of
post--AGB SEDs.
The LI sources (\#3\& \#4 in \Fg\SED ) are clearly of ``class I''.
Of the five class--I sources in van der Veen \etal (1989),
three are located in IRAS regions IV and VIII
to the left of the evolutionary sequence.
%
The RI sources (\#5 -- \#10 in \Fg\SED ) are all ``class IVa''. 
According to van der Veen \etal (1989), there is a large
difference in initial mass between
the two classes : 4 \msun\ for class I and 1.7 \msun\ for
class IVa (from luminosity--distance self--consistency arguments).
This is in excellent agreement with our derived
mass range of 1--4 \msun\ for the RI sources and the fact
that the LI sources could be the more massive (higher mass loss)
sources (\Sct 3.1).
Other properties of the post--AGB groups are in Table \SPEC ;
both classes have a high percentage of O--rich objects.
The class--I sources, hence the LI stars, would still have
considerable mass loss (van der Veen \etal1989).

\section{Precursors and successors}

Since the LI sources have higher outflow velocities than
the RI sources,
the first assumption is that they have higher luminosities
and/or metallicity (van der Veen 1989).
Clearly, we already found indications that the LI sources
are more massive in \Sct 3.1\&4 .
Our observations could be explained, if the
``early'' transitions of the HOW97 models (\Sct 3.2) were to happen 
only in the more massive objects.

In an attempt to trace further intrinsic differences between 
the two groups of stars, we first determined the average
latitude of the 25 LI sources, \decdeg0.46, and 
of the 31 RI sources, \decdeg1.66 (\Fg\LAT, Table \RILI ).
Average latitude, or angular vertical scale, is directly proportional to the
absolute vertical scale, $h_{\rm z}$, of a sample
and inversely to average distance. Itself, $h_{\rm z}$ 
is a function of the age $t$ of the sample, via the vertical
velocity dispersion $\sigma_{\rm w}$. 
The dispersion increases as $\sigma_{\rm w}\, \propto \, t^{\sim 0.5}$
(Wielen 1977, for stars from 0.2 Gyr to 5 Gyr; Just \etal1996).
Close to the plane, $h_{\rm z}\, \propto \, \sigma_{\rm w}$ and
out of the plane
$h_{\rm z}\, \propto \, \sigma^2_{\rm w}$  (Just \etal 1996).
For the LI group, $h_{\rm z}\,\propto\, t^{0.5}$ is 
definitely appropriate (\Fg\LAT ).
For the RI sample, reaching out to latitudes well over 3\degr\ (\Fg\LAT ),
the exponential is possible larger than 0.5 (and smaller than 1).
All in all, the LI sources
are either much younger than the RI sources or much further away.

\begin{deluxetable}{crrrrrrrrrrr}
\tablecaption{Measured averages and model values ($M_{\rm i}$
as assumed; columns
5,6 for AGB--tip and $Z$=0.02 from Bertelli \etal1994).}
\tablewidth{0pt}
\tablehead{
\colhead{ } &
\colhead{$<V_e>$ } &
\colhead{$<b>$ } &
\colhead{$ <S_{12,25,60}>$ } &
\colhead{ } &
\colhead{$M_i$ } &
\colhead{t } &
\colhead{$L_\ast$ } &
}
\startdata
 & km/s & $^\circ$ & Jy & & \msun & Gyr & \lsun   \\
RI& 13.1& 1.66 &4,25,29& & 1.7 & 1.8& 1.0E4 \\
LI& 17.5& 0.46 &28,68,88& & 4.0 & 0.2& 3.2E4 \\
\enddata
\end{deluxetable}

Second, the stellar luminosities should
be roughly proportional to $V_{\rm exp}^4$ 
for similar metallicities :

\eqnam\ELV
$$ L_{\ast} \propto V_{\rm exp}^4 Z^{-2 }, 
\eqno(\new) $$

\noindent
following van der Veen (1989).
So, $L_{\ast}^{\rm LI} \sim 3.2 L_{\ast}^{\rm RI}$ (Table \RILI). 
This clearly fits
in well with the previous argument that the LI sources should be
much younger than the RI sources, if at similar distances.

\begin{deluxetable}{crrrrrrrrrrr}
\tablecaption{Measured and predicted ratios (LI/RI)}
\tablewidth{0pt}
\tablehead{
\colhead{ } &
\colhead{$L_\ast $ } &
\colhead{$<b>$ } &
\colhead{$S_{\rm OH}$  }
}
\startdata
predicted & $\sim$3.0 & \lsim 0.33 &  $\sim$0.4 \\
measured & 3.3$\pm$0.1 & \lsim 0.28 &  $\sim$0.5 \\
\enddata
\end{deluxetable}

Third, the average bolometric correction $BC_{12}$ at 12 \mum\ for the
RI sources is 5.3, for the LI sources 2.7 (van der Veen \& Breukers 1989).
The ratio of the total bolometric fluxes then would be
$S_{\rm bol}^{\rm LI}/S_{\rm bol}^{\rm RI}=
 (S_{12}*BC_{12})^{\rm LI}/ (S_{12}*BC_{12})^{\rm RI}\sim$ 6.75/2 = 3.4 
on average.

From arguments 1+2, we can assume
that the two groups are at similar distances, and from
arguments 2+3 that they have similar metallicities.
Hence, from the second argument, 
LI stars are three times more luminous than the RI stars
and indeed originated from more massive main sequence stars.

\begin{deluxetable}{crrrrrrrrrrr}
\tablecaption{Details for the LI sources. 
The $S_{\rm OH}$ is for the highest peak.}
\tablewidth{0pt}
\tablehead{
\colhead{ Name (OH$\ell,b$) } &
\colhead{ IRAS } &
\colhead{ $V_{\rm exp}\rm ({km\over s})$ } &
\colhead{ $S_{\rm OH}$(Jy) } &
\colhead{  } &
\colhead{ } &
\colhead{ }
}
\startdata
OH326.518$-$00.633 & 15452$-$5459&  9.5 &  9.27& & & \\
OH328.225$+$00.042 & 15514$-$5323& 19.7 & 57.78& & & \\
OH344.929$+$00.014 & 17004$-$4119 & 19.0 & 39.93& & & \\
OH351.592$+$00.318 & 17193$-$3546 & 10.2 &  0.78& & & \\
OH351.607$+$00.022 & 17205$-$3556& 18.3 &  0.33 & & & \\
OH351.118$-$00.352 & 17207$-$3632& 16.8 &  1.36 & & & \\
OH353.637$+$00.815 & 17230$-$3348&  9.5 &  2.77 & & & \\
OH354.642$+$00.830 & 17256$-$3258& 32.8 &  0.58 & & & \\
OH358.425$-$00.175 & 17392$-$3020& 21.2 &  1.00 & & & \\
OH357.988$-$00.988 & 17414$-$3108& 14.6 &  0.55 & & & \\
OH010.076$-$00.095 & 18052$-$2016& 28.5 &  2.02 & & & \\
OH007.452$-$02.615 & 18092$-$2347& 17.0 &  0.20  & & & \\
OH011.522$-$00.582 & 18100$-$1915& 13.6 &  1.10 & & & \\
OH012.973$+$00.133 & 18103$-$1738& 21.5 &  0.80 & & & \\
OH016.117$-$00.291 & 18182$-$1504& 21.5 &  9.40 & & & \\
OH020.679$+$00.084 & 18257$-$1052& 18.2 &  5.80 & & & \\
OH021.457$+$00.491 & 18257$-$1000& 18.1 & 18.90 & & & \\
OH022.993$-$00.285 & 18314$-$0900& 14.8 &  8.30 & & & \\
OH024.692$+$00.235 & 18327$-$0715& 19.3 &  4.40 & & & \\
OH025.495$-$00.288 & 18361$-$0647& 15.9 &  1.80 & & & \\
OH030.715$+$00.427 & 18432$-$0149& 17.0 &  3.90 & & & \\
OH030.554$+$00.281 & 18434$-$0202& 19.2 &  1.10 & & & \\
OH030.091$-$00.686 & 18460$-$0254& 19.2 & 35.10 & & & \\
OH031.984$-$00.485 & 18488$-$0107& 19.3 &  5.10 & & & \\
OH038.101$-$00.125 & 18588$+$0428& 21.5 &  0.40 & & & \\
\enddata
\end{deluxetable}

So, let's assume that the LI sources had average initial
masses of 4\msun\  and the RI sources 1.7\msun (see Table \SPEC ).
Corresponding AGB--tip luminosities and ages (Bertelli \etal 1994)
are in Table \RILI .
The luminosities are a factor of two higher than those given 
in Table \SPEC . The latter are current, post--AGB luminosities,
whereas the former are the luminosities the objects would have
had on the tip of the AGB. Post--AGB luminosities are also 
given by Vassiliadis \& Wood (1994) and interpolate to
$\sim$6350\lsun\ and $\sim$17000\lsun\ for those masses, 
both for H--burning and solar metallicity.

Importantly, the luminosity ratio between the two groups is
3$^{+0.2}_{-0.3}$ from all three references, compared to 3.2 from the
outflow velocities or 3.4 from the bolometric fluxes.
For $<b> \propto t^{>0.5}$, the ratio of 
average latitudes would be \lsim 0.33, compared 
to \lsim 0.28 from the observations.
For the assumed initial masses, van
der Veen (1989) gives AGB outflow velocities 
of 13\kms\ and 17\kms , respectively.
All observed values are closely reproduced (Table \RILI).
The median OH--peak flux density
for the RI sample is twice that for the LI sample.
This is close to what is expected from 
$S_{\rm OH} \propto S_{\rm bol}^{1.5} X_{21}^2$
(Paper I, \Eqt 3b), giving a ratio of 
${3.4^{1.5}}*{0.25^2}\sim$0.4 .

From a typical initial--mass function, we would expect
about 10 times more sources between 1--4\msun\ than between
4--6\msun\ (the AGB limit).
For our OH sample, however, the numbers are roughly equal.
OH selection effects do not explain this discrepancy,
since the OH luminosities of the RI sample are
higher than those of the LI sample, if indeed the average distances 
to the two samples are the same, as argued above.
From the model tracks by HOW97, however (see \Sct 3.2),
the LI sources are expected to take much longer than
the RI sources to reach an effective temperature of 6500 K after
leaving the AGB (see \Sct 3.2).
Since the effective temperature plays an important role in 
the termination of the OH masers, this effect could make up the numbers.

\begin{deluxetable}{crrrrrrrrrrr}
\tablecaption{Details for the RI sources.
The $S_{\rm OH}$ is for the highest peak.}
\tablewidth{0pt}
\tablehead{
\colhead{ Name (OH$\ell,b$) } &
\colhead{ IRAS } &
\colhead{ $V_{\rm exp}\rm ({km\over s})$ } &
\colhead{ $S_{\rm OH}$(Jy) } &
\colhead{  } &
\colhead{ } &
\colhead{ }
}
\startdata
OH314.933$-$02.052 & 14341$-$6211&  9.5 &  0.64 & & & \\
OH335.832$+$01.434 & 16209$-$4714& 13.9 &  1.23& & & \\
OH338.507$-$02.915 & 16507$-$4810& 11.7 &  0.62& & & \\
OH349.949$+$01.537 & 17097$-$3624& 12.4 &  0.70& & & \\
OH353.973$+$02.727 & 17164$-$3226& 13.9 &  2.33& & & \\
OH349.804$-$00.321 & 17168$-$3736& 13.1 & 10.89& & & \\
OH348.813$-$02.840 & 17245$-$3951& 11.7 & 13.09& & & \\
OH353.945$-$00.972 & 17310$-$3432& 11.7 &  1.39& & & \\
OH359.750$+$02.629 & 17317$-$2743& 11.0 &  4.60& & & \\
OH359.140$+$01.137 & 17359$-$2902&  9.5 &  1.66& & & \\
OH355.111$-$01.697 & 17370$-$3357& 11.7 &  2.77& & & \\
OH355.641$-$01.742 & 17385$-$3332& 10.2 &  4.03& & & \\
OH000.892$+$01.342 & 17393$-$2727& 14.6 & 47.87& & & \\
OH001.212$+$01.257 & 17404$-$2713& 13.1 &  4.48& & & \\
OH359.233$-$01.876 & 17479$-$3032& 13.1 &  1.94& & & \\
OH004.007$+$00.915 & 17482$-$2501& 14.6 &  1.25& & & \\
OH000.072$-$02.044 & 17506$-$2955& 10.9 &  0.76& & & \\
OH002.286$-$01.801 & 17548$-$2753& 16.1 &  1.42& & & \\
OH007.961$+$01.445 & 17550$-$2120& 12.4 &  4.30 & & & \\
OH008.854$+$01.689 & 17560$-$2027& 13.6 &  3.00& & & \\
OH004.017$-$01.679 & 17582$-$2619& 11.0 &  1.02& & & \\
OH006.594$-$02.011 & 18051$-$2415& 13.6 &  0.40& & & \\
OH015.364$+$01.925 & 18087$-$1440& 14.8 &  3.40& & & \\
OH015.700$+$00.770 & 18135$-$1456& 14.8 &  9.60& & & \\
OH017.684$-$02.032 & 18276$-$1431& 11.4 & 39.00 & & & \\
OH025.057$-$00.350 &18355$-$0712 & 11.4 &  2.80& & & \\
OH027.577$-$00.853 & 18420$-$0512& 12.4 &  3.00& & & \\
OH030.394$-$00.706 & 18467$-$0238& 27.2 &  0.60& & & \\
OH038.909$+$03.178 & 18485$+$0642& 13.6 & 11.50& & & \\
OH037.118$-$00.847 & 18596$+$0315& 13.6 &  6.60& & & \\
OH035.209$-$02.653 & 19024$+$0044& 13.6 &  4.00& & & \\
\enddata
\end{deluxetable}

The 60--\mum\ flux--density measurements of the LI stars
might be overestimated in the IRAS point--source
catalogue due to their very low latitudes. However, using
the available flags in the IRAS PSC, both the LI and the
RI sources are found to be mostly in confused regions.
The CIRR1 and CIRR2 flags average 6 to 7 for both 
groups and CIRR3 is 254 for all RI and LI sources (IRAS Explanatory
Supplement).
Thus, as mentioned in the introduction, it is likely that the
60--\mum\ fluxes are overestimated for all sources in our
low--latitude survey, but the LI sources are not necessarily
suffering more from this effect.

A mechanism that may cause different evolution
at the tip of the AGB for different initial
masses could be hot--bottom burning. Only AGB stars more massive
than $\sim$4\msun\ undergo this process. This mass limit fits
in very well with what we have found in this paper.
In \Fg\IRSD , an object is plotted that is argued to be
hot--bottom burning (van Loon \etal 2001). 
It is IRAS 05298$-$6957, an LMC cluster OH/IR star.
This object is indeed found to have the properties of an LI star.
Its OH outflow velocity is 11\kms\ (Wood \etal1992) which is consistent
with typical LI luminosity, for $Z$=0.008 instead of 0.02.

\subsection{Massive precursors to bipolar PN ?}

It is interesting to revisit the scaleheights of the two groups (\Fg\LAT ).
Following Sevenster (1999), the apparent scaleheights deproject to
roughly 100 pc and 300 pc, respectively. Due to our narrow 
latitude coverage, the latter value may be somewhat underestimated.
Those values are close to
what van der Veen found for the two spectral classes 
(Table \SPEC ) and correspond closely to
the scaleheights given by Corradi (2000)
for bipolar (130 pc) and elliptical (320 pc) planetary nebulae.
Also, the location of the ``bipolar outflow'' region 
defined by Zijlstra \etal(2001) suggests that the LI sources
may be related to bipolar sources.
The higher fraction of irregular
spectra amongst LI sources and the preferred location of 
sources with irregular spectra in this ``bipolar outflow'' region,
as found in Paper I, indicate that on their way ``back'' 
to the red PN region of the IRAS two--colour diagram the LI 
sources may go through a highly irregular, high--outflow mass--loss phase.
The bipolar planetary nebulae themselves might again be located
to the right of the evolutionary sequence (\Fg\IRSE ; HOW97), but
no longer harbour OH masers.
One would expect the average $R_{32}$ of bipolar PNe to be higher
than that of elliptical PNe, if this scenario were correct.
This appears to be the case : Corradi \& Schwarz (1995) show
that for $R_{21}$ > 2 bipolars are located at 0 < $R_{32}$ < 2 and
ellipticals at -1 < $R_{32}$ < 1, the same range as the RI sources.

\section{Conclusions}

A red and a blue group of OH--masing post--AGB stars are
discussed in detail. 
The consistency of a variety of independent 
arguments provides a solid basis for the assumption that
the blue sources are indeed post--AGB objects;
their existence was predicted by van Hoof \etal(1997).
Using evolutionary--sequence turn--off colours, 
spectral energy distributions, outflow velocities and 
scaleheight, we argue that the two groups separate by
initial mass at $M_{\rm i}\,=\,4\,$\msun . 

The higher--mass objects
may have AGB termination earlier (at still longer periods)
than the lower--mass objects
and make a blue--$R_{21}$ loop through the IRAS two--colour diagram.
They have a 15--\mum\ excess, as well as strong 60,100--\mum\ excess.
The lower--mass objects show very sudden reddening
at $R_{21}$$\sim$1.2, which 
corresponds to average initial masses of $\sim$1.7\msun .
These sources must all undergo
a very similar process at AGB termination.

At the end of their OH--masing post--AGB stages, the 
high--mass, blue objects may
go through an irregular mass--loss phase (Paper I), preceding the
bipolar planetary--nebulae stage. The low--mass, red objects probably
evolve into elliptical planetary nebulae.

The combined colour $ 2.5\log((S^{\rm msx}_{21}\,S^{\rm iras}_{12})/
(S^{\rm msx}_{8}\,S^{\rm iras}_{25}))$ proves very sensitive to
distinguish lower--mass, early post--AGB objects from sources still
on the AGB and also to distinguish more evolved post--AGB objects from
star--forming regions. 


\begin{acknowledgements}

MS thanks Peter van Hoof for making the model tracks available and
offering useful comments about this paper. The Leidse Sterrewacht
kindly provided
desk and computer to finish the final version of this paper.

\end{acknowledgements}

{}

\onecolumn
\vfill
\eject

\begin{figure}
\fignam\IRSB
\anfig
\psfig{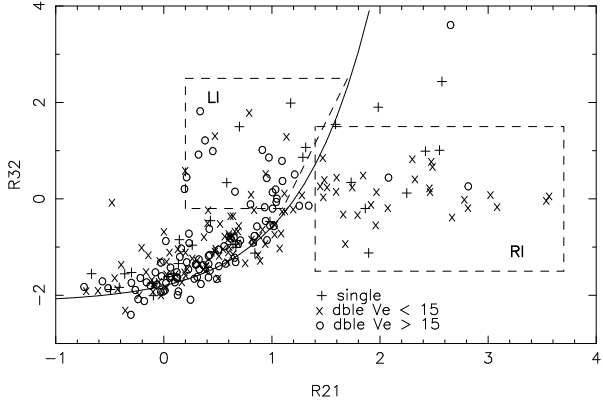}
\figcaption{
The IRAS two--colour diagram with symbols according to outflow
velocity.  The ``standard'' post--AGB region  to the right
of the evolutionary sequence (``RI'' sources) is
populated mostly by sources with outflow velocities below 15\kms .
In the region to the left, on the other hand,
sources have mostly higher outflow velocities (``LI'' sources). 
}
\end{figure}
\vfill
\eject

\begin{figure}
\fignam\IRSD
\anfig
\psfig{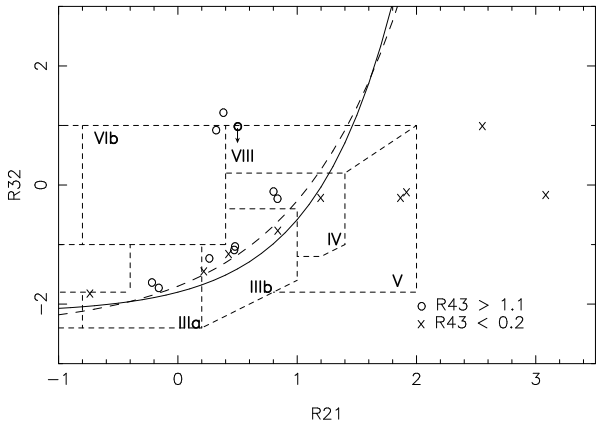}
\figcaption{
The IRAS two--colour diagram showing the sources selected
by $R_{43}$ ($[60-100]$).  Only very few sources have well--defined
$S_{100}$, but the ``LI'' and ``RI'' (\Sct 2)
branches clearly stand out, in an exaggerated way, in this selection.
Be aware that the 100--\mum\ flux--density measurements are most likely to 
be overestimated for the LI sources, as they are close
to the Plane (see \Sct 5).
The fat circle with arrow indicates the hot--bottom--burning
OH/IR star in the LMC (van Loon \etal2001) 
discussed in \Sct 5.
The solid curve is the theoretical evolutionary sequence (\Eqt\EST ),
the dashed curve is a common observational evolutionary sequence
($R_{32} = -2.42 + 0.72\,\exp[1.1\,R_{21}] $).
The dashed boxes are the regions as defined in Van der Veen \& Habing (1988).
}
\end{figure}
\vfill
\eject

\begin{figure*}
\fignam\MSXD
\anfig
\psfig{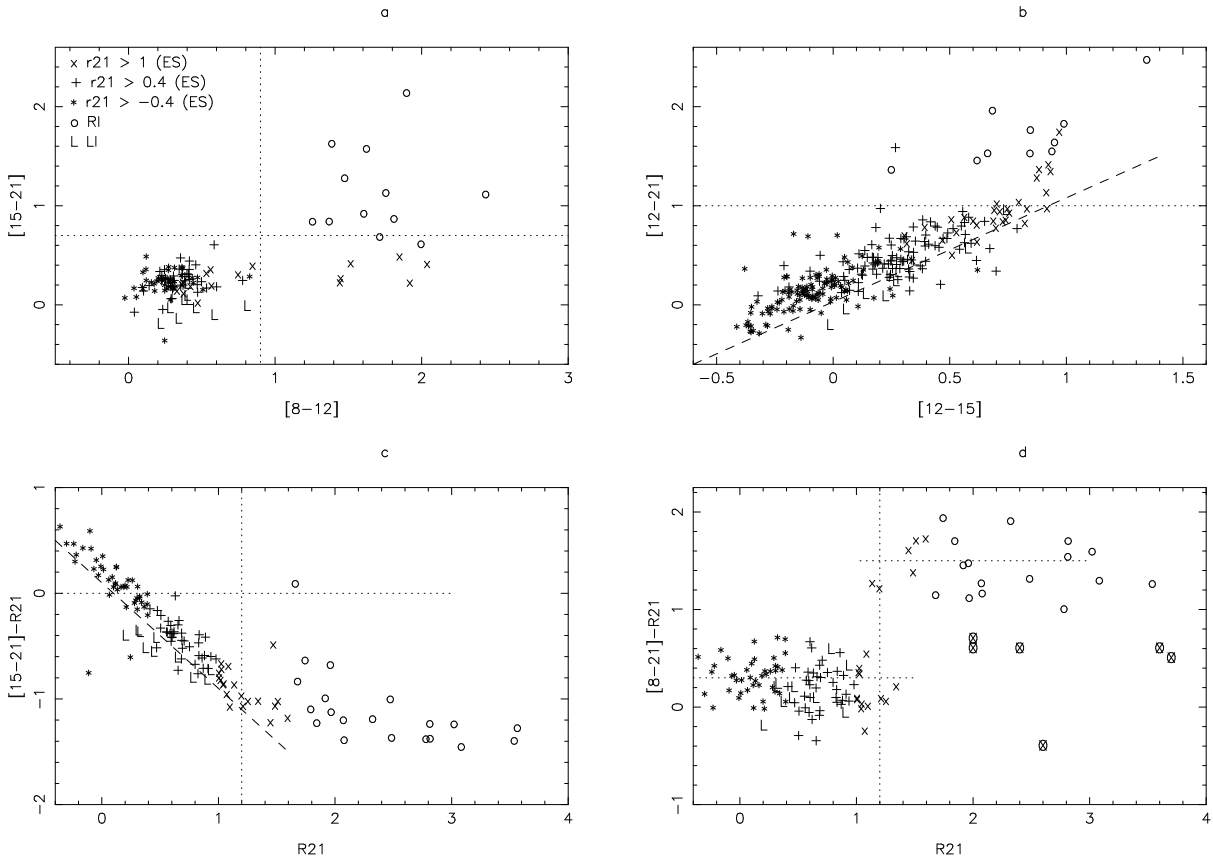}
\figcaption{
The top panels ({\bf a,b}) show two MSX two--colour diagrams
and the bottom panels ({\bf c,d}) plot the IRAS colour $R_{21}$ versus
the difference between MSX colours and $R_{21}$.
In all panels, only double--OH--peaked sources are plotted,
with symbols according to IRAS colours :
($\ast$,+,x) for sources close to the evolutionary
sequence with increasing $R_{21}$, as well as the
LI (`L')and RI (`o') sources as defined in \Sct 2. 
The LI sources clearly have the bluest $[15-21]$ of all and
slightly redder $[12-15]$ than the bulk of the AGB sources.
The MSX colour $[8-21]$ is the best equivalent of $R_{21}$.
They are almost equal for $R_{21}$$\,<\,$1.2 ($T$=170 K)
and $[8-21]$ is four times redder for $R_{21}$$\,>\,$1.2.
The transition is very sudden transition and 
corresponds to an average initial mass of 1.7 \msun\ (\Sct 3.1).
In plot {\bf d}, some known OH--masing star--forming regions 
(Caswell 1998,1999) are plotted for reference (encircled crosses).
Dashed and dotted lines in all panels are plotted to guide the eye.
This figure is discussed in detail in \Sct 3.1.
}
\end{figure*}

\vfill
\eject

\begin{figure}
\fignam\IRSE
\anfig
\psfig{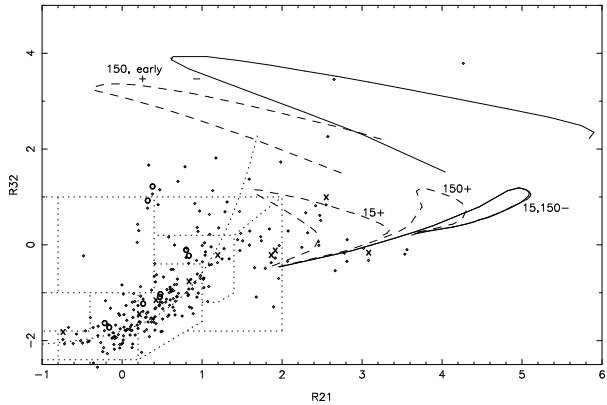}
\figcaption{
The IRAS two--colour diagram, with all objects indicated by dots and
circles and crosses as \Fg\IRSD .
Overlaid are 3--\msun\ tracks from HOW97, for post--AGB wind of 15 \kms\ and
150 \kms\ with (dashed) and without (solid) post--AGB dustformation.
For 150 \kms , there are also the models for early AGB termination.
A ``15+(early)'' model would possibly 
reach in to the regions where the LI sources are found.
}
\end{figure}
\vfill
\eject

\begin{figure}
\fignam\HFC
\anfig
\psfig{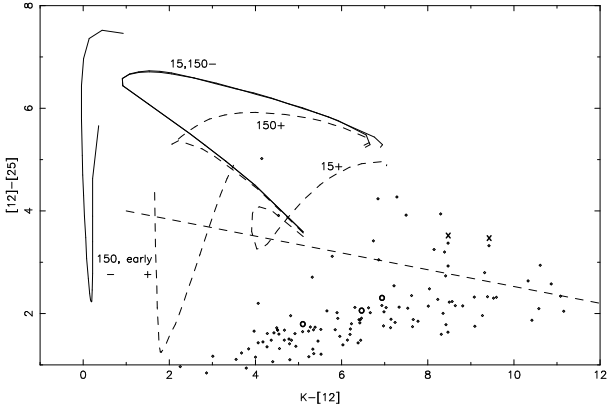}
\figcaption{
The mid--near--infrared colour diagram as advocated by HOW97, 
with tracks as in \Fg\IRSE .
Circles indicate LI sources, crosses are sources 
with $R_{43}<$0.2 (\Fg\IRSD ).
Note that the circles and crosses are not the same selection as
in \Fg\IRSE , but do indicate similar sources.
The dashed line line separates the RI sources from the 
(bluer) AGB sources. The tracks are calculated for a 
distance of 1 kpc. As the sources are
mostly well further than that, they can be dereddened by
moving them to lower K$-[12]$ (hardly any reddening for [12]$-$[25]).
Again, a ``15+(early)'' model might be the best candidate to reproduce
the (blue) colours of the LI sources.
}
\end{figure}
\vfill
\eject

\begin{figure*}
\fignam\SED
\anfig
\psfig{figure=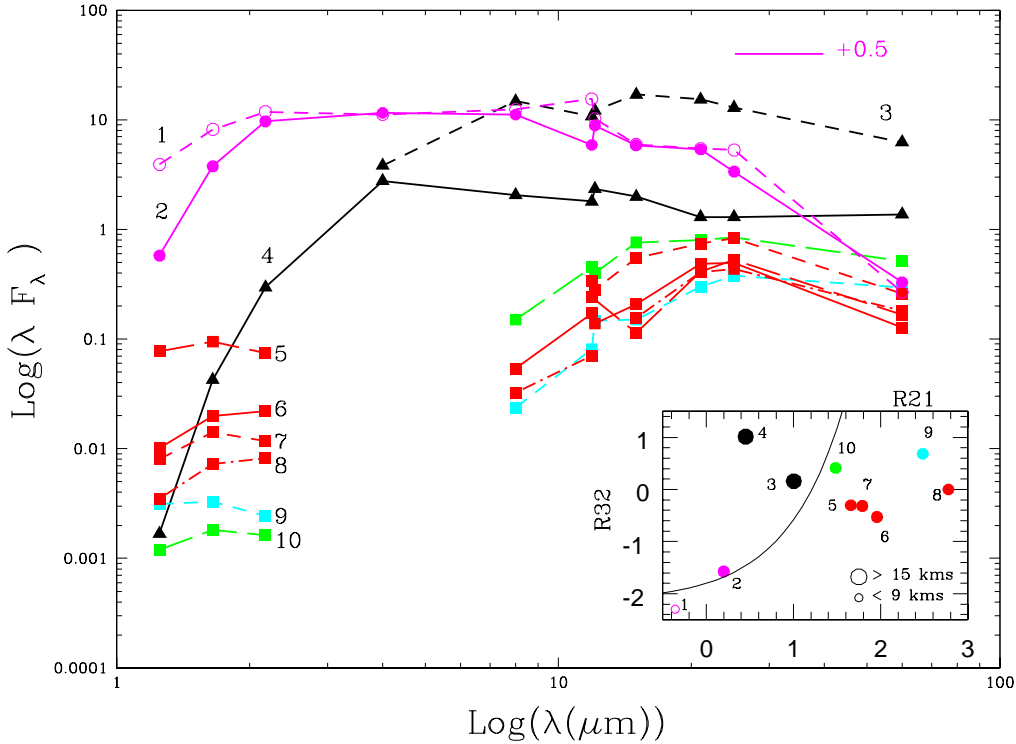,width=14truecm}
\figcaption{
The spectral energy distributions composed of the 
flux densities in the NIR bands (2MASS) and the MIR bands
(MSX, IRAS) for 2 AGB sources (\#1\&\#2,cyan), 2 LI sources (\#3\&\#4,black)
and 6 RI sources (\#5 --\#10,red,green,turquoise).
The IRAS 12--\mum\ flux densities, including the 9.7--\mum\ silicate feature,
is plotted slightly to the left of the MSX 12--\mum\ measurement, that
does not include 9.7 \mum .
The difference between the two may give an indication of the 
optical depth in the silicate feature.
The inserted plot shows the location of the sources in the IRAS two--colour
diagram, with symbol sizes scaled according to outflow velocity :
smallest is $V_{\rm exp}$$\,<\,$9 \kms , largest 
is $V_{\rm exp}$$\,>\,$15 \kms\ and
intermediate is in between. The open symbol in the insert
corresponds to the SED with open symbols in the main plot.
For the RI sources, the MSX 4--\mum\ flux 
densities are undetermined and some other points are missing in several
of the SEDs. The AGB SEDs (\#1\&\#2, 
cyan) are shifted upward by half a magnitude
to avoid confusing overlap with the LI SED. The vertical axis is in
arbitrary units.
}
\end{figure*}
\vfill
\eject

\begin{figure}
\fignam\LAT
\anfig
\psfig{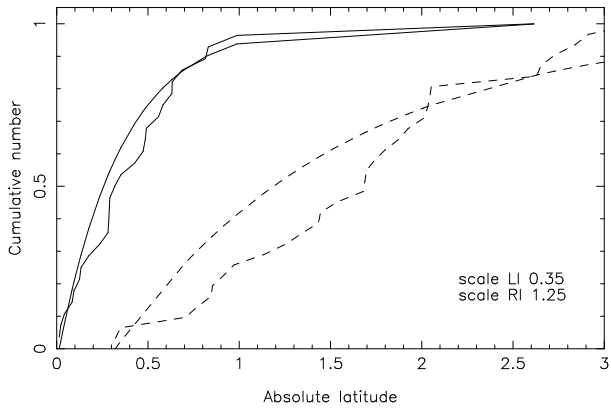}
\figcaption{
The cumulative distributions versus absolute latitude
for the LI (solid) and RI (dashed) stars.
The Kolmogorov--Smirnov probability that the two distributions are the
same is 3$\times 10^{-9}$.
The smooth fits are exponentials with apparent scaleheights \decdeg 0.35 
(100 pc) and \decdeg 1.25 (300 pc), respectively (see \Sct 5.1 and
compare to Table \SPEC ).
}
\end{figure}

\end{document}